\documentclass[aps,prl,twocolumn,superscriptaddress,showpacs]{revtex4-1}

\usepackage[colorlinks,linkcolor=blue,citecolor=blue,urlcolor=blue,breaklinks=true]{hyperref}
\usepackage{amsmath,amssymb,graphics,epsfig,epstopdf,color,verbatim,braket,tabularx,breakurl}

\begin{document}

\title{Interaction-Induced Characteristic Length in Strongly Many-Body Localized Systems}

\author{Rong-Qiang He}
\email{rqhe@ruc.edu.cn}
\affiliation{Institute for Advanced Study, Tsinghua University, Beijing 100084, China}
\affiliation{Department of Physics, Renmin University of China, Beijing 100872, China}
\author{Zhong-Yi Lu}
\email{zlu@ruc.edu.cn}
\affiliation{Department of Physics, Renmin University of China, Beijing 100872, China}

\date{\today}

\begin{abstract}
  We propose a numerical method for explicitly constructing a complete set of local integrals of motion (LIOM) and definitely show the existence of LIOM for strongly many-body localized systems. The method combines exact diagonalization and nonlinear minimization, and gradually deforms the LIOM for the noninteracting case to those for the interacting case. By using this method we find that for strongly disordered and weakly interacting systems, there are two characteristic lengths in the LIOM. The first one is governed by disorder and is of Anderson-localization nature. The second one is induced by interaction but shows a discontinuity at zero interaction, showing a nonperturbative nature. We prove that the entanglement and correlation in any eigenstate extend not longer than twice the second length and thus the eigenstates of the system are `quasi-product states' with such a localization length.

\end{abstract}

\pacs{71.23.An, 72.15.Rn, 02.60.Pn, 05.30.-d}

\maketitle

A variety of theoretical \cite{Basko06,Gornyi05,Imbrie16,Imbrie16JSP} and numerical \cite{Oganesyan07,Znidaric08,Pal10,Cecile10,Canovi11,Kjall14,Luitz15,Tang15} studies have shown that disorder induced localization is robust against small interactions, known as many-body localization (MBL). Various remarkable new phenomena have been found or conceived for MBL systems, such as area law entanglement entropy of eigenstates \cite{Bauer13}, logarithmic divergence of entanglement entropy in time and space \cite{Bardarson12,Serbyn13slow,Kim14}, identically zero conductance \cite{Ros15}, ergodicity violation, breakdown of the eigenstate thermalization hypothesis (ETH) \cite{Deutsch91,Srednicki94,Tasaki98,Rigol07,Rigol08} and quantum statistical mechanics, protection for quantum and topological orders at finite temperatures \cite{Wootton11,Stark11,Huse13,Bauer13,Bahri13}. Most of them can be readily derived from the concept of an extensive set of local integrals of motion (LIOM), which is argued to exist in MBL systems.

Conceptually, LIOM are independent (quasi-)local operators commuting with each other as well as the Hamiltonian, and determine the basic structure of an MBL system since formally the Hamiltonian can be clearly expressed in terms of a complete set of LIOM \cite{Serbyn13,Huse14}. However, in practice, it is difficult to find them explicitly and accurately. Ros {\it et al.} \cite{Ros15} and Imbrie \cite{Imbrie16,Imbrie16JSP} used perturbation theories. Rademaker and Ortu\~no \cite{Rademaker16} employed displacement operators. They eliminated off-diagonal interaction terms in the Hamiltonian order by order, but stopped halfway because of exponential increase of computational cost, which results in a truncation error. Chandran {\it et al.} \cite{Chandran15} used long-time evolution of local operators. Inglis and Pollet \cite{Inglis16} found incomplete and approximate LIOM with nonlinear minimization.

In this Letter, we propose a numerical method for constructing a complete set of LIOM explicitly and accurately for strongly disordered MBL systems. The method starts with the LIOM for the noninteracting case (which can be obtained readily and exactly), deforms them into the LIOM for the interacting case via consecutive unitary transformations. The locality of the LIOM is preserved in the transformations. Our method is so accurate that we can extract localization lengths hidden in the exponentially decaying tails of the LIOM. We identify a characteristic localization length induced by interaction, which is longer than the one induced solely by disorder. This implies that interaction tends to delocalize the system as we expect. It is however remarkable that the characteristic length induced by interaction shows a discontinuity at zero interaction, suggesting that interaction has a nonperturbative effect on disordered systems. Furthermore, we prove that the length induced by interaction characterizes the localization of eigenstates of MBL systems and beyond twice of this characteristic length no entanglement or correlation can survive.

{\em Model.}---As an example, we study a spinless fermionic model on a one-dimensional chain with nearest-neighbor hoppings, nearest-neighbor density-density interactions, and disordered on-site potentials. The Hamiltonian is
\begin{equation}\label{eq:model}
\hat{H} = -t \sum_{\langle ij\rangle} (c_{i}^{\dagger}c_{j}+c_{j}^{\dagger}c_{i}) + V \sum_{\langle ij\rangle} \hat{n}_{i} \hat{n}_{j} - \sum_{i} \mu_i \hat{n}_i.
\end{equation}
Here we set $t = 1$. $\mu_i$ is randomly chosen in $[-w, w]$ with uniform distribution. $L$ is the number of lattice sites. Open boundary condition is chosen in the calculation. For $V = 0$, the system is known to be Anderson localized \cite{Anderson58}, while for large $w$ and small $V$, the system is MBL. We would like to emphasize that the applications of the following concepts and discussions are not restricted to this specific model.

{\em LIOM.}---A set of LIOM for a system is a set of independent and (quasi-)local operators $\{\hat{J}^{(i)}\}$ commuting with each other as well as the Hamiltonian, i.e., $[\hat{J}^{(i)}, \hat{J}^{(j)}] = [\hat{J}^{(i)}, \hat{H}] = 0$. A simple example is the one for model (\ref{eq:model}) of the noninteracting case ($V = 0$), denoted by $\{\hat{J}_0^{(i)}\}$, which consists of the occupancy number operators of the localized single-particle eigenstates.  More specifically, $\hat{J}_0^{(i)} = d_i^\dagger d_i = \hat{Q} \hat{N}_0^{(i)} \hat{Q}^\dagger$ and $d_i^\dagger \equiv \hat{Q} c_i^\dagger \hat{Q}^\dagger$, where $\hat{N}_0^{(i)} \equiv \hat{n}_i$ and $d_i^\dagger$ creates a particle in a localized single-particle eigenstate and $\hat{Q}$ is a unitary operator diagonalizing the noninteracting Hamiltonian in the Fock basis (the eigenstates of $\hat{n}_i$), i.e, $\mathbf{H}_0 \equiv \mathbf{Q}^\dagger \mathbf{H} \mathbf{Q}$ is diagonal for $V = 0$. (Remind: $\mathbf{H}_0$ is {\em not} diagonal for $V \neq 0$.) The matrix representation of an operator $\hat{O}$ in the Fock basis is denoted as $\mathbf{O}$.

{\em Constructing LIOM.}---We can connect the interacting case $\{\hat{J}^{(i)}\}$ to the noninteracting case $\{\hat{J}_0^{(i)}\}$ by a unitary transformation $\hat{U}$: $\hat{J}^{(i)} = \hat{Q} \hat{U} \hat{Q}^\dagger \hat{J}_0^{(i)} \hat{Q} \hat{U}^\dagger \hat{Q}^\dagger$, which reduces to the LIOM operator for the noninteracting case $\hat{J}_0^{(i)}$ by setting $\hat{U}$ to the identity operator $\hat{I}$. Obviously, $[\hat{J}^{(i)}, \hat{J}^{(j)}] = 0$. The condition $[\hat{J}^{(i)}, \hat{H}] = 0$ requires that $[\hat{N}_0^{(i)}, \hat{U}^\dagger \hat{H}_0 \hat{U}] = 0$, equivalently,
\begin{equation}\label{eq:aim}
[\mathbf{N}_0^{(i)}, \mathbf{U}^\dagger \mathbf{H}_0 \mathbf{U}] = 0.
\end{equation}
Meanwhile, we want $\hat{J}^{(i)}$ to be as localized as possible. For a weak interaction, this can be realized by tailoring $\mathbf{U}$ to be {\em as close to $\mathbf{I}$ as possible} (ACTIAP) by noting that $\hat{J}_0^{(i)}$ is localized and $\hat{J}^{(i)}$ is most close to $\hat{J}_0^{(i)}$ when $\mathbf{U}$ is most close to $\mathbf{I}$. On the other hand, a strong interaction will destroy the MBL and the LIOM will no longer exist. To the end, we want to find a $\mathbf{U}$ such that Eq.~(\ref{eq:aim}) is satisfied and meanwhile ACTIAP.

{\em Algorithm 1} (partial determination of $\mathbf{U}$): Eq.~(\ref{eq:aim}) is satisfied if $\mathbf{U}^\dagger \mathbf{H}_0 \mathbf{U}$ is diagonal because $\mathbf{N}_0^{(i)}$ is diagonal and two diagonal matrices commute. So we may want to construct $\hat{U}$ by diagonalizing $\mathbf{H}_0$, and $\mathbf{U}$ is composed of the eigenvectors (as its columns). Next, we try to order the eigenvectors so that $\mathbf{U}$ is ACTIAP, which is possible when $\mathbf{H}_0$ is diagonally dominant. We refer to the absolute value of the largest element of an eigenvector as the `principle value' of the eigenvector. We try to order the eigenvectors so that their largest elements in absolute value is located on the diagonal of $\mathbf{U}$. Specifically, the positions of the eigenvectors in $\mathbf{U}$ are determined one by one in descending order of their principle values. Sometimes, there may be a problem in this process: the correct position of an eigenvector coincides with that of another eigenvector determined previously and the process has to be halted. This means only a part of $\mathbf{U}$ is determined. The undetermined part has to be dealt with further. After a suitable column permutation $\mathbf{P}$, the determined part of $\mathbf{U}$ can be grouped to the left side (denoted as $\mathbf{U}^{\rm d}$) and the undetermined to the right (denoted as $\mathbf{U}^{\rm u}$), i.e., $\mathbf{U} \mathbf{P} = [\mathbf{U}^{\rm d} | \mathbf{U}^{\rm u}]$. Set $\mathbf{U}^{\rm u} \equiv \mathbf{R} \mathbf{T}$, where $\mathbf{R}$ is determined so that $[\mathbf{U}^{\rm d} | \mathbf{R}] \mathbf{P}^\dagger$ is unitary and ACTIAP \footnote{$\mathbf{R} = f(g(\mathbf{E}))$. Every element of $\mathbf{E}$ is 0 or 1. It is 1 only when it is permuted via $\mathbf{P}^\dagger$ to a diagonal element of $[\mathbf{U}^{\rm d} | \mathbf{E}] \mathbf{P}^\dagger$. Function $g(\mathbf{E})$ orthonomalizes every column of $\mathbf{E}$ to all columns of $\mathbf{U}^{\rm d}$. Function $f(\mathbf{S})$ unitarizes $\mathbf{S}$ in the space spanned by the columns of $\mathbf{S}$ via repeated applications of $\mathbf{S} \leftarrow \frac{3}{2} \mathbf{S} - \frac{1}{2} \mathbf{S} \mathbf{S}^\dagger \mathbf{S}$ until convergence, i.e., $\mathbf{S} = \frac{3}{2} \mathbf{S} - \frac{1}{2} \mathbf{S} \mathbf{S}^\dagger \mathbf{S}$. In other words, function $f(\mathbf{S})$ produces a matrix close to $\mathbf{S}$ and so that $[\mathbf{U}^{\rm d} | \mathbf{R}]$ is unitary.}, and $\mathbf{T}$ is a unitary matrix with smaller dimension than that of $\mathbf{U}$ and is to be determined and have to be ACTIAP. Now, Eq.~(\ref{eq:aim}) reduces to
\begin{equation}\label{eq:aim1p}
[\mathbf{N}_1^{(i)}, \mathbf{T}^\dagger \mathbf{H}_1^{\prime} \mathbf{T}] = 0,
\end{equation}
where $\mathbf{H}_1^{\prime} \equiv \mathbf{R}^\dagger \mathbf{H}_0 \mathbf{R}$ and $\mathbf{N}_1^{(i)}$ (being diagonal) is the lower-right block of $\mathbf{P}^\dagger \mathbf{N}_0^{(i)} \mathbf{P}$.

{\em Algorithm 2} (improving diagonal dominance): Few-small-step \footnote{The number of steps in this minimization is 1 in this work, which is found to be most efficient for the whole LIOM construction. The step size $|\delta \mathbf{X}|$ is set to $|\delta \mathbf{X}| / |\mathbf{I}| \sim 0.08$, to which the minimization efficiency is not very sensitive.} steepest-descent minimization of $\chi(\mathbf{X}) \equiv \sum_i {\rm tr} [ \mathbf{K}^{(i)\dagger} \mathbf{K}^{(i)} ]$, beginning with $\mathbf{X} = 0$, where $\mathbf{K}^{(i)} \equiv [\mathbf{N}_1^{(i)}, \exp(\mathbf{X}^\dagger) \mathbf{H}_1^{\prime} \exp(\mathbf{X})]$ with $\mathbf{X}^\dagger = - \mathbf{X}$. After minimization we obtain $\mathbf{X}$, which is small. $\exp(\mathbf{X})$ is unitary and is close to $\mathbf{I}$. Set $\mathbf{T} = \exp(\mathbf{X}) \mathbf{U}_1$. Now, Eq.~(\ref{eq:aim1p}) reduces to
\begin{equation}\label{eq:aim1}
[\mathbf{N}_1^{(i)}, \mathbf{U}_1^\dagger \mathbf{H}_1 \mathbf{U}_1] = 0,
\end{equation}
where $\mathbf{H}_1 \equiv \mathbf{U}_1^\dagger \mathbf{H}_1^{\prime} \mathbf{U}_1$ is more diagonally dominant than $\mathbf{H}_1^{\prime}$ and $\mathbf{U}_1$ is to be determined and have to be ACTIAP.

Now, by comparing Eq.~(\ref{eq:aim1}) with Eq.~(\ref{eq:aim}), we find that we are in a situation resembling to the beginning. Therefore, we can apply algorithms 1 and 2 to Eq.~(\ref{eq:aim1}) with $\mathbf{N}_1^{(i)}$, $\mathbf{H}_1$, and $\mathbf{U}_1$ instead of $\mathbf{N}_0^{(i)}$, $\mathbf{H}_0$, and $\mathbf{U}$, respectively. Further, we iterate this process until $\mathbf{U}$ is fully determined. In practice, we find that only a few iterations will accomplish the LIOM construction. Remark: this LIOM construction algorithm is an improved version of that in Ref.~\cite{He2016liom} and is much more efficient.

Finding $\hat{U}$ explicitly is a merit of this method. The raising and lowering operators transforming the simultaneous eigenstates of $\{\hat{J}^{(i)}\}$ and $\hat{H}$ from one to another can be readily obtained by
\begin{equation}\label{eq:cd}
\hat{J}_+^{(i)} \equiv \hat{Q} \hat{U} c_i^\dagger \hat{U}^\dagger \hat{Q}^\dagger, \quad \hat{J}_-^{(i)} \equiv \hat{J}_+^{(i)\dagger}.
\end{equation}
$\hat{J}_+^{(i)}$ ($\hat{J}_-^{(i)}$) raises (lowers) the eigenvalues of $\hat{J}^{(i)}$. By the way, $\hat{J}^{(i)} = \hat{J}_+^{(i)} \hat{J}_-^{(i)}$.

{\em Locality of operators.}---To explicitly characterize locality, we introduce a reduction error $\eta$ for reducing an operator $O$ into a region $A$ as
\begin{equation}\label{eq:locality}
{\eta_A}(O) \equiv \sqrt {\frac{{\rm tr} (O - \tilde O)^\dagger (O - \tilde O)}{{\rm tr} {\tilde O}^ \dagger \tilde O}}, \quad \tilde O \equiv \frac{{\rm tr}_B O}{{\rm tr}_B I},
\end{equation}
where $B$ denotes the region excluding $A$ and $I$ is the identity operator. $A$ is called the target region. $\tilde O$ is the reduced operator of $O$. $O$ is reducible for a target region $A$ if and only if (1) $O = \tilde O$, or (2) $O = O_A \otimes I_B = O_A$, or (3) $O$ is totally localized in $A$, where $O_A$ is some operator defined in region $A$ and $I_B$ is the identity operator in region $B$. If $O = O_A \otimes O_B$ with $O_B = \lambda (I_B + X_B)$ and $X_B$ is small and traceless (where $\lambda$ is a constant),
\begin{equation}\label{eq:otilde}
\tilde O = O_A \otimes (\lambda I_B), \quad {\eta_A}(O) = \sqrt{\frac{{\rm tr}_B X_B^\dagger X_B}{{\rm tr}_B I_B^ \dagger I_B}}.
\end{equation}
Thus, smaller ${\eta_A}(O)$ means that $O$ is more localized in $A$. In other words, ${\eta_A}(O)$ quantitatively measures the localization of $O$ in region $A$.

\begin{figure}[b]
  \includegraphics[width=\columnwidth]{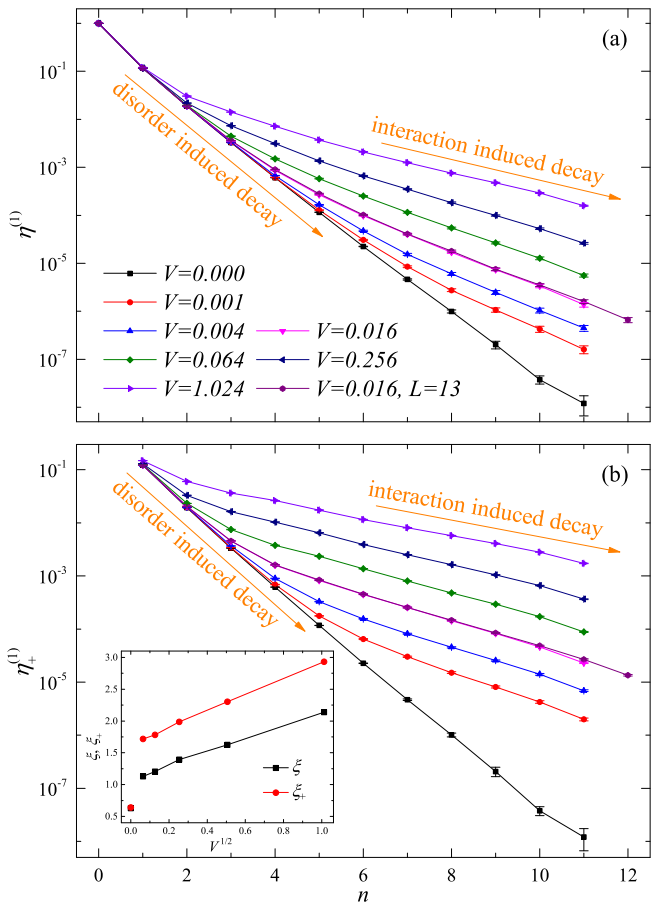}
  \caption{\label{fig:etaxi} (Color online) Disorder averaged reduction errors of (a) the LIOM operator $\hat{J}^{(1)}$ and (b) the corresponding raising operator $\hat{J}_+^{(1)}$ localized near one chain end. $\eta^{(1)}(n) \equiv \eta_n(\hat{J}^{(1)})$ and $\eta_+^{(1)}(n) \equiv \eta_n(\hat{J}_+^{(1)})$, where the target region includes the $n$ sites starting from that chain end. $w = 32$. $L = 12$, except the ones explicitly labeled ($L = 13$). The exponential decay of the reduction errors have only one phase for $V = 0$, which is disorder induced, while the corresponding ones have two phases for $V > 0$, featuring the disorder induced decay for small $n$ and the interaction induced decay for large $n$, respectively. The inset in (b) shows the characteristic lengths extracted from the exponential decays. It is evident that a discontinuity appears at $V = 0^+$ for both $\xi$ and $\xi_+$.}
\end{figure}

{\em Results.}---We focus on $\eta^{(1)}(n) \equiv \eta_n(\hat{J}^{(1)})$ and $\eta_+^{(1)}(n) \equiv \eta_n(\hat{J}_+^{(1)})$, the reduction errors of $\hat{J}^{(1)}$ and its corresponding raising operator $\hat{J}_+^{(1)}$, where $\hat{J}^{(1)}$ is the LIOM operator at one chain end and the target region for reduction includes the $n$ sites starting from that chain end. $\eta_n(\hat{J}_+^{(1)}) \equiv \eta_n(\hat{J}_-^{(1)})$. Because $\hat{J}^{(1)}$ localizes near the chain end, $\eta^{(1)}(n)$ and $\eta_+^{(1)}(n)$ decrease as $n$ increases. Particularly, $\eta^{(1)}(0) = 1$ and $\eta^{(1)}(L) = \eta_+^{(1)}(L) = 0$. Note that $\eta_+^{(1)}(0)$ is not well defined because ${\rm tr} \hat{J}_+^{(1)} = 0$.

The disorder averaged reduction errors $\eta^{(1)}(n)$ and $\eta_+^{(1)}(n)$ are plotted in Fig.~\ref{fig:etaxi}. The average is taken over $N_d(V)$ independent disorder realizations. Specifically, $5 \times 10^{-6} N_d = 1, 2, 5, 10, 20, 50, 100$ for $10^3 V = 1024, 256, 64, 16, 4, 1, 0$, respectively. As the interaction decreases, the fluctuation of the reduction error $\delta \eta^{(1)} / \eta^{(1)}$ or $\delta \eta_+^{(1)} / \eta_+^{(1)}$ tends to be divergent for large $n$ and therefore we need a very large number of disorder realizations to make the statistic error small enough for small interactions.

It is clearly seen in Fig.~\ref{fig:etaxi} that the reduction errors decay exponentially, indicating that the LIOM and the raising/lowering operators are localized in the system for both the noninteracting ($V = 0$) and interacting ($V > 0$) cases. However, there are two remarkable differences between the two cases. First, $\eta^{(1)} = \eta_+^{(1)}$ for $V = 0$ but $\eta^{(1)} < \eta_+^{(1)}$ for $V > 0$. This shows that for the interacting case $\hat{J}_+^{(1)}$ is less localized than $\hat{J}^{(1)}$, i.e., changing the eigenvalue of the LIOM operator $\hat{J}^{(1)}$ entails changing the state outside the support of $\hat{J}^{(1)}$, suggesting that the eigenstates of the system is entangled.

Second, the exponential decay of $\eta^{(1)}(n)$ or $\eta_+^{(1)}(n)$ has only one phase for $V = 0$, which is disorder induced. However, there are two phases for $V > 0$. The first phase appears for small $n$, where the decay of the reduction error follows the one in the noninteracting case and thus is disorder driven. The second phase appears for large $n$, where the decay of the reduction error is evidently slower than that in the noninteracting case and thus is interaction driven.

The exponentially decaying tails of the reduction errors can be well modelled by
\begin{equation}\label{eq:xi}
\eta(n) = \eta(0) \exp(-n/\xi),
\end{equation}
where $\xi$ defines a characteristic localization length, and the coefficient $\eta(0)$ indicates the amplitude. For $\xi$ and $\eta(0)$, we add subscript `$+$' for those corresponding to the raising operator $\hat{J}_+^{(1)}$. $\xi$ and $\xi_+$ are shown in the inset of Fig.~\ref{fig:etaxi}(b). To get rid of the finite size effect, we choose a sufficiently large disorder strength ($w = 32$) so that the MBL length is much smaller than the system size ($L = 12$). The disappearing of the finite size effect can be seen by comparing the $L = 12$ and $L = 13$ results in Fig.~\ref{fig:etaxi}. The differences of the reduction errors at $n = 10$ and $11$ result from a boundary effect; these two points are not used to calculate $\xi$ and $\xi_+$.

As shown in the inset of Fig.~\ref{fig:etaxi}(b), $\xi$ and $\xi_+$ increase as $V$ increases. This means that the interaction tends to delocalize the LIOM operators and hence the system, as what we expect. However, it is surprising that there are discontinuities in $\xi$ and $\xi_+$ as $V \rightarrow 0^+$. This means that an arbitrarily small interaction can induce a localization length longer than the one induced by disorder and change the system qualitatively. This suggests that interactions have a nonperturbative effect on disordered systems. It has been known that an arbitrarily small interaction can result in a logarithmic growth of entanglement entropy in time \cite{Serbyn13slow}. So it is interesting to raise such a question how the interaction induced localization length affects the dynamics of an MBL system.

{\em Quasi-product states.}---As we show above, $\xi < \xi_+$. A natural question is which length is more relevant to MBL? Let us first consider the roles of their corresponding operators. The eigenvalues of the LIOM operators uniquely label the eigenstates of the Hamiltonian while the raising/lowering operators transform the eigenstates. For an eigenstate $| \psi \rangle$, $\hat{K}^{(i)} | \psi \rangle$ is another eigenstate, where
\begin{equation}\label{eq:qps}
\hat{K}^{(i)} \equiv \hat{J}_+^{(i)} + \hat{J}_-^{(i)}
\end{equation}
is hermitian and unitary. Intuitively, the two eigenstates differ only in the support of $\hat{J}_+^{(i)}$ or $\hat{J}_-^{(i)}$, which suggests that $\xi_+$ characterizes the length scale determining the structure of the eigenstates.

\begin{figure}[b]
  \includegraphics[width=\columnwidth]{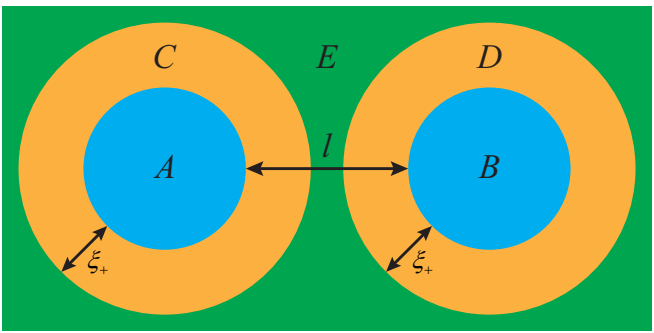}
  \caption{\label{fig:qps} (Color online) A two-dimensional illustration of short-range entanglement of an eigenstate of an MBL system. If the separation distance $l$ between regions $A$ and $B$ is much larger than twice the characteristic length $\xi_+$ of the raising/lowering operators, there will be no entanglement or correlation between regions $A$ and $B$ because region $E$ can be unitarily transformed into vacuum without affecting regions $A$ and $B$.
  }
\end{figure}

Actually, we can prove that the entanglement and correlation of an eigenstate only extend to a length scale characterized by $2 \xi_+$, beyond which there is no entanglement or correlation. See Fig.~\ref{fig:qps}. For $l \gg 2\xi_+$, any eigenstate $| \psi \rangle$ can be written as
\begin{equation}\label{eq:qps}\nonumber
| \psi \rangle = \prod_{i} \hat{K}^{(i)} \prod_{j} \hat{K}^{(j)} \prod_{k} \hat{K}^{(k)} | {\rm vac} \rangle = \prod_{i} \hat{K}^{(i)} | \psi \prime \rangle,
\end{equation}
where $| {\rm vac} \rangle$ is the vacuum, $| \psi \prime \rangle \equiv \prod_{j} \hat{K}^{(j)} \prod_{k} \hat{K}^{(k)} | {\rm vac} \rangle$, $\hat{K}^{(j)}$ and $\hat{K}^{(k)}$ have their supports in region $A + C$ and region $B + D$, respectively, while the support of $\hat{K}^{(i)}$ may be in region $E$ or overlap with region $C$ and/or region $D$ but never overlap with region $A$ or $B$. So region $E$ is vacuum in $| \psi \prime \rangle$, and $| \psi \prime \rangle$ is a product state of the two states in regions $A + C$ and $B + D$, respectively, which results in
\begin{equation}\label{eq:productstate}
\hat{\rho}\prime_{A+C+B+D} = \hat{\rho}\prime_{A+C} \otimes \hat{\rho}\prime_{B+D},
\end{equation}
where $\hat{\rho}\prime_{A+C+B+D}$, $\hat{\rho}\prime_{A+C}$, and $\hat{\rho}\prime_{B+D}$ are the reduced density operators, for example, $\hat{\rho}\prime_{A+C} = {\rm tr}_{B+D+E} \hat{\rho}\prime$ with $\hat{\rho}\prime \equiv | \psi \prime \rangle \langle \psi \prime |$. Evidently, $| \psi \prime \rangle$ has no entanglement or correlation between regions $A$ and $B$, i.e., the corresponding reduced density operators satisfy
\begin{equation}\label{eq:ent}
\hat{\rho}\prime_{A+B} = \hat{\rho}\prime_{A} \otimes \hat{\rho}\prime_{B},
\end{equation}
which, actually, can be derived by acting ${\rm tr}_{C+D}$ on both the sides of Eq.~(\ref{eq:productstate}). By using the fact that ${\rm tr}_{S} \hat{U^\dagger} \hat{O} \hat{U} = {\rm tr}_{S} \hat{O} \hat{U} \hat{U^\dagger} = {\rm tr}_{S} \hat{O}$ for any operator $\hat{O}$ and region $S$ if $\hat{U}$ is a unitary operator and has its support in region $S$, we can, finally, obtain
\begin{equation}\label{eq:entcor}
\hat{\rho}_{A+B} = \hat{\rho}_{A} \otimes \hat{\rho}_{B}
\end{equation}
with $\hat{\rho} \equiv | \psi \rangle \langle \psi  |$ by noting that $\hat{\rho}_{A} = \hat{\rho}\prime_{A}$, $\hat{\rho}_{B} = \hat{\rho}\prime_{B}$, and $\hat{\rho}_{A+B} = \hat{\rho}\prime_{A+B}$ since $\prod_{i} \hat{K}^{(i)}$, connecting $\hat{\rho}$ with $\hat{\rho}\prime$, is unitary and its support does not overlap with region $A$ or $B$. Q.E.D.

Due to the existence of interaction, generally, an eigenstate of an MBL system cannot be factorized as a product state like the case of a noninteracting Anderson localized system, but it looks very much like, because any two spatially $2\xi_+$-separated points in the system have no entanglement or correlation as we proved above. We call it a `quasi-product state' (QPS). Evidently, every eigenstate of an MBL system is a QPS if $\xi_+$ is finite. $\xi_+$ plays a similar role as the localization length of single-particle states of an Anderson-localized system and hence can be considered as the eigenstate localization length of an MBL system.

{\em Conclusion.}---We have proposed a numerical method for explicitly and accurately constructing a complete set of LIOM, and shown definitely the existence of LIOM in strongly MBL systems. A one-dimensional disordered spinless fermionic model with nearest-neighbor interactions has been studied as a role model. By inspecting the reduction errors of the LIOM and raising/lowering operators, we have identified a characteristic length ($\xi_+$) induced by interaction, which decreases as the interaction decreases, but shows a discontinuity at zero interaction, i.e., an infinitesimal interaction will induce a characteristic length longer than the one induced solely by disorder. This reveals that interaction has a nonperturbative impact on disordered systems. We have proved that beyond a distance of $2 \xi_+$ there is no entanglement or correlation in any eigenstate of the system. So the eigenstates of the system are `quasi-product states' with a localization length of $\xi_+$.

\begin{acknowledgments}
This work was supported by National Natural Science Foundation of China (Grants No. 11474356 and No. 91421304).  R.Q.H. was supported by China Postdoctoral Science Foundation (Grant No. 2015T80069). Computational resources were provided by National Supercomputer Center in Guangzhou with Tianhe-2 Supercomputer and Physical Laboratory of High Performance Computing in RUC.
\end{acknowledgments}

\bibliography{ref}

\end{document}